\documentclass[amssymb,prb,twocolumn,showpacs,superscriptaddress]{revtex4}
\usepackage{epsfig}
\usepackage{dcolumn}
\usepackage{amsmath,amssymb}
\DeclareMathOperator{\var}{var}

\begin{document}

\title{Thermoelectric transport of mesoscopic conductors coupled to voltage and thermal probes} 
\author{David S\'anchez}
\affiliation{Institut de F\'{\i}sica Interdisciplin\`aria i de Sistemes Complexos
IFISC (CSIC-UIB), E-07122 Palma de Mallorca, Spain}
\affiliation{Departament de F\'{\i}sica,
Universitat de les Illes Balears, E-07122 Palma de Mallorca, Spain}
\author{Lloren\c{c} Serra}
\affiliation{Institut de F\'{\i}sica Interdisciplin\`aria i de Sistemes Complexos
IFISC (CSIC-UIB), E-07122 Palma de Mallorca, Spain}
\affiliation{Departament de F\'{\i}sica,
Universitat de les Illes Balears, E-07122 Palma de Mallorca, Spain}
\date{\today}

\begin{abstract}
We investigate basic properties of the thermopower (Seebeck 
coefficient) of phase-coherent conductors
under the influence of dephasing and inelastic processes.
Transport across the system is caused by a voltage
bias or a thermal gradient applied between two terminals.
Inelastic scattering is modeled with the aid of an additional probe acting
as an ideal potentiometer and thermometer.
We find that inelastic scattering reduces the conductor's thermopower
and, more unexpectedly, generates a magnetic-field asymmetry
in the Seebeck coefficient. The latter effect is shown to be
a higher-order effect in the Sommerfeld expansion.
We discuss our result  using two illustrative examples.
First, we consider a generic mesoscopic system
described within random matrix theory and demonstrate that
thermopower fluctuations disappear quickly as the number of probe modes increases. 
Second, the asymmetry is explicitly calculated in the quantum limit
of a ballistic microjunction. We find that asymmetric scattering
strongly enhances the effect and discuss its dependence on temperature
and Fermi energy.
\end{abstract}

\maketitle

{\em Introduction}. Recent advances in heat measurements
have enabled to envisage promising thermoelectric applications at the mesoscale. \cite{gia06}
In particular, there have been a number of exciting proposals ranging
from thermal analogs of electronic rectifiers \cite{ter02} and transistors \cite{wan07}
to efficient converters of heat into electricity \cite{bel08} which can be shown
to reach optimized configurations. \cite{san10} Of fundamental importance is the
experimental verification of the carriers' charge sign using thermopower techniques only. \cite{red07}

Since these novel devices operate at the quantum regime of transport,
it is of primary interest to investigate dephasing effects which may be detrimental
to their performance. Additionally, energy flow at finite temperature
is expected to be particularly sensitive to 
inelastic transitions inside the sample due, e.g., to coupling to phonons.
This problem has been addressed only very recently. \cite{lei10,ent10}

A convenient way to introduce dephasing and energy relaxation in a mesoscopic
conductor is based on the voltage probe model. \cite{but86} A fictitious terminal
is attached to the sample such that the net current
flowing through the probe is set to zero. Hence, a carrier absorbed by the probe
is reemitted into the conductor with an unrelated phase.
The clear advantage of this approach lies on 
its simplicity and its independence of the microscopic details of the
phase-randomizing processes. Therefore, universal properties of generic conductors
can be investigated this way. \cite{bar95,bro97} The model has been recently
applied to find the temperature and chemical potential profiles
of an array of quantum dots. \cite{jac09}

Including a third terminal acting as a dephasing probe requires to extend
the scattering approach to the multiterminal case. For thermoelectric transport,
this was achieved by Butcher. \cite{but90}
Experimentally, there are already available data for paradigmatic mesoscopic
systems such as two-terminal point contacts \cite{mol92} and chaotic cavities. \cite{god99}
The multiterminal case, however, has been less explored.
A very recent exception is Ref. \onlinecite{mat11} where a voltage drop is generated
transversally to a horizontal thermal gradient in a four-terminal setup
with an asymmetric scatterer.

\begin{figure}[t]
\centerline{
\epsfig{file=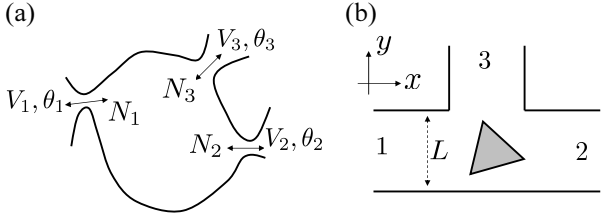,angle=0,width=0.45\textwidth,clip}
}
\caption{Sketches of (a) a chaotic cavity attached to voltage and thermal reservoirs
and (b) a microjunction with an asymmetric scatterer coupled to a probe as a
model of a quantum wire with incoherent scattering.}
\label{fig1}
\end{figure}

Here we investigate incoherent scattering effects on the thermopower
(Seebeck coefficient $S$),
defined at linear response as the ratio of voltage $\Delta V$ to temperature
$\Delta\theta$ differences applied between two terminals (1 and 2 in Fig.\ \ref{fig1})
at vanishing current $I$,
\begin{equation}
\label{eqtp}
S \equiv 
\left.
\frac{\Delta V}{\Delta\theta}
\right|_{I=0}\; ,
\end{equation}
in the presence of an attached probe (terminal 3 in Fig.\ \ref{fig1})
acting as an ideal potentiometer {\em and} thermometer.
Quite generally, we find that $S$ decreases as compared to the case without probe
and, strikingly, the presence of incoherent scattering due to the probe
causes the development of a magnetic-field asymmetry in $S$.
To illustrate our findings, we analyze
(i) a mesoscopic conductor with generic properties (a chaotic cavity); and
(ii) a ballistic wire in the quantum limit (a few propagating modes)
with an asymmetric scatterer. In the latter case, we explicitly calculate
the size of the magnetoasymmetry. This result is relevant in view of recent predictions
that relate this asymmetry to efficiency limits of
thermal devices in converting a temperature gradient
into electrical work. \cite{ben11,sai11}

{\em Model}. We consider a mesoscopic conductor coupled to three terminals.
Thermoelectric transport is described by the electric current, $I_\alpha$,
and the heat current, $Q_\alpha$, flowing from terminal $\alpha=1,2,3$.
In the linear response regime, the transport equations read: 
\begin{align}\label{eq_iq}
I_\alpha &= \sum_\beta G_{\alpha\beta} V_\beta+\sum_\beta L_{\alpha\beta} \theta_\beta \,,\\
Q_\alpha &=\sum_\beta M_{\alpha\beta} V_\beta+\sum_\beta K_{\alpha\beta} \theta_\beta \,,
\end{align}
where $eV_\alpha=E_F-\mu_\alpha$ are voltage shifts away from the common Fermi energy $E_F$,
with $\mu_\alpha$ the electrochemical potential of lead $\alpha$,
while $\theta_\alpha=\theta-T_\alpha$ measures
deviations of the reservoir temperature $T_\alpha$ from the common (bath) temperature $\theta$.
The transport coefficients are expressed
in terms of the transmission probability $T_{\alpha\beta}$:
\begin{align}
\label{eq_rc1}
G_{\alpha\beta} &= -g_V \int\! dE  \,(N_\alpha \,\delta_{\alpha\beta}-T_{\alpha\beta}) f_0'\,,\\
L_{\alpha\beta} &= -\frac{g_V}{e\theta} \int\!dE \,(E-E_F)(N_\alpha\,\delta_{\alpha\beta}-T_{\alpha\beta})f_0'\label{eq_rc2}\,,\\
M_{\alpha\beta}&=-\theta L_{\alpha\beta}\,,\label{eq_rc3}\\
K_{\alpha\beta} &= \frac{g_V}{e^2\theta} \int\!dE\, (E-E_F)^2 (N_\alpha\,\delta_{\alpha\beta}-T_{\alpha\beta})f_0'\label{eq_rc4}\,,
\end{align}
with $g_V=2e^ 2/h$ the conductance quantum, $N_\alpha$ the number of
propagating channels in lead $\alpha$ and $f_0'$
the energy derivative of the Fermi distribution function evaluated at $V_\alpha=\theta_\alpha=0$.
In Eq.~\eqref{eq_rc3} we have used the Onsager reciprocity relations between cross terms.

{\em Isothermal case}. We first consider the case where the temperature
probe is an externally fixed parameter. This would correspond, e.g., to a phonon
subsystem maintained at a different temperature. \cite{ent10} The probe thus works as
an ideal voltmeter with voltage $V_3$ determined from the condition $I_3=0$,
\begin{align}\label{eq_v3iso}
V_3&=\frac{G_{31} V_1+G_{32}V_2}{G_{31}+G_{32}}\nonumber\\
&+\frac{L_{31}}{G_{31}+G_{32}}(\theta_1-\theta_3)+
\frac{L_{32}}{G_{31}+G_{32}}(\theta_2-\theta_3)\,.
\end{align}
We substitute Eq.~\eqref{eq_v3iso} into the expression for $I_1$,
from which we find the current
$I=I_1=-I_2$ flowing through the system: 
\begin{equation}\label{eq_minusi}
-I=g V_{12}+L_{12}\theta_{12}+L_{13}\theta_{13}
-\frac{G_{13}}{G_{31}+G_{32}}[L_{31}\theta_{13}+L_{32}\theta_{23}]\,.
\end{equation}
$I$ is a function of voltage
$V_{ij}=V_i-V_j$ ($i,j=\{1,2,3\}$) and temperature
$\theta_{ij}=\theta_i-\theta_j$
differences, as should be. In Eq.~\eqref{eq_minusi},
$-g=G_{12}+G_{13}G_{32}/(G_{31}+G_{32})$
is the well-known expression
for the conductance in the presence of a voltage probe
in the purely electric case for which all temperature
shifts are set to zero. \cite{but86}

In the isothermal configuration, the probe is maintained at the same temperature
as the common bath ($\theta_3=0$). We are then free to specify the precise form of the temperature
gradient $\theta_1-\theta_2\neq 0$. We choose the symmetric bias
$\theta_1=-\theta_2=\Delta\theta/2$, which is commonly employed in actual measurements. \cite{mat11}
We then compute the thermopower from Eq.\ (\ref{eqtp})  for $V_1-V_2=\Delta V$:
\begin{equation}\label{eq_S}
S=\frac{1}{g} \left[ L_{12}+\frac{1}{2}L_{13}+\frac{1}{2}\frac{G_{13}}{G_{31}+G_{32}}(L_{32}-L_{31})\right]\,.
\end{equation}

Equation~\eqref{eq_S} shows two contributions to the thermopower as compared
to the case without phase randomization for which $S=-L_{12}/G_{12}$.
The first part corresponds to direct inelastic scattering
with the probe and is proportional to the term $L_{13}$, as can be expected from an analogy
with the purely electric case (cf., $g$ above). The second
part is more surprising---it is nonzero only if there is an asymmetry between the transmissions
into the probe of those carriers injected from lead 1 and lead 2 ($L_{32}\neq L_{31}$).
This term has no counterpart in the purely electric case. A similar asymmetry effect,
but with phonon carriers, has been recently shown to be crucial
in the development of rectification effects in dielectric junctions. \cite{min10}

{\em Adiabatic case}. More interesting is the case when the probe
plays simultaneously the role of an ideal potentiometer {\em and} thermometer.
Then, not only the charge current but also the energy flux is zero at the probe.
These two conditions determine $V_3$ and $\theta_3$. \cite{eng81}
Recently, it has been suggested
that electronic local temperatures can be consistently defined
introducing thermal probes. \cite{cas10}

Imposing $I_3=0$ and $Q_3=0$, we find:
\begin{align}
V_{32} &=-\frac{(G_{31}K_{33} +\theta L_{31}L_{33})V_{12}+(L_{31}K_{33}-K_{31}L_{33})\theta_{12}}
{G_{33}K_{33}+\theta L_{33}^2} \,,\\
\theta_{32} &=-\frac{(G_{33}K_{31} + \theta L_{31}L_{33})\theta_{12}+(G_{31}L_{33}-L_{31}G_{33})\theta V_{12}}
{G_{33}K_{33}+\theta L_{33}^2}\,,
\end{align}
which we substitute in the equation $I_1=0$ in order to obtain the
thermopower $S=\Delta V/\Delta\theta=V_{12}/\theta_{12}$,
\begin{multline}\label{eq_sad}
S=\left[G_{13} (L_{31}K_{33} -K_{31} L_{33})+G_{33} (L_{13} K_{31}-L_{11} K_{33})\right. \\
\left. +\theta L_{33} (L_{13} L_{31} -L_{11} L_{33})\right]
/\left[\theta L_{33}(G_{11} L_{33}-G_{31} L_{13})\right. \\
\left.+G_{33} (\theta L_{13} L_{31} +G_{11}K_{33})
-G_{13} (G_{31} K_{33}+\theta L_{31} L_{33})\right]\,.
\end{multline}

{\em Chaotic cavity}. We now consider a generic mesoscopic sample---a metallic quantum dot
whose classical analog displays chaotic dynamics [see Fig.~\ref{fig1}(a)].
Its isotropic properties permits us to treat the cavity as an effectively zero-dimensional
object characterized with a mean level spacing $\delta$.
Transport occurs when the cavity is coupled to external reservoirs usually
via quantum point contacts with a large number ($N_1$ and $N_2$)
of propagating channels. The experimentally relevant case deals with
clean samples. \cite{mar97} As a consequence, transport is ballistic
and its corresponding statistics
is well described by random matrix theory (RMT). \cite{bar94}
In what follows, we assume $eV_\alpha,k_B \theta_\alpha< E_T$,
with $E_T= (N_1+N_2)\delta$ the Thouless energy.

Using a Sommerfeld expansion, one finds approximate expressions for the
response coefficients in Eqs.~(\ref{eq_rc1})--(\ref{eq_rc4})
in terms of the transmission functions
$T_{\alpha\beta}$ and their energy derivatives
$T_{\alpha\beta}'$ evaluated at the Fermi level:
$G_{\alpha\beta} =g_V (N_\alpha\delta_{\alpha\beta}-T_{\alpha\beta})$,
$L_{\alpha\beta} =g_\theta T_{\alpha\beta}'$ and
$K_{\alpha\beta} =-g_\theta (N_\alpha\delta_{\alpha\beta}-T_{\alpha\beta})/e$,
where $g_\theta=(2e/h) (\pi^ 2k_B^2\theta/3)$.
Note that the ratio $\theta g_V/eg_\theta=\pi^ 2k_B^2/3$ depends on universal constants
only, representing the fundamental quantum of thermal conductance of a perfectly
transmitting mode with averaged reservoir temperature $\theta$. \cite{shw00}

We investigate the statistical properties of the thermopower,
in particular its mean value and variance. 
In the limit $N_\alpha \gg 1$ we can calculate correlations of $T_{\alpha\beta}$
and $T_{\alpha\beta}'$ within RMT. We consider the cases $\beta=1$ (orthogonal ensemble)
and $\beta=2$ (unitary ensemble) corresponding, respectively, to the presence and the absence
of time reversal symmetry. 
We define the functions $A_{\mu\nu}=N_\mu N_\nu /N_t^2-\delta_{\mu\nu}N_\mu/N_t$
and $B_{\rho\sigma}=\sqrt{2}(N_\rho N_\sigma /N_t^3-\delta_{\rho\sigma}N_\rho/N_t^2)$,
where $\mu,\nu,\rho,\sigma$ are lead indices and $N_t$ is the total number of modes.
Then, using Ref. \onlinecite{pol03} we derive the useful relations,
\begin{align}
\langle T_ {\mu\nu}\rangle &= \delta_{\mu\nu}N_\mu - N_\mu N_\nu /N_t+\delta_{\beta1}A_{\mu\nu}\,,\\
\langle T_ {\mu\nu}' \rangle &=0\label{eq_G0}\,,\\
\langle T_ {\mu\nu} T_ {\rho\sigma} \rangle  &= A_{\mu\rho}A_ {\nu\sigma}
+\delta_{\beta1}A_{\mu\sigma}A_ {\nu\rho}\,,\\
\langle T_ {\mu\nu}' T_ {\rho\sigma}' \rangle  &= B_{\mu\rho}B_ {\nu\sigma}
+\delta_{\beta1}B_{\mu\sigma}B_ {\nu\rho}\,.
\end{align}

From Eq. ~\eqref{eq_G0} we immediately see that the mean vanishes, $\langle S\rangle=0$.
This result was found in Ref.~\onlinecite{lan98} for the case without dephasing. The reason is clear:
the $L$ coefficients are functions of the energy derivative of the scattering matrix,
but this is zero on average (the derivative can be positive or negative for a specific sample
of the ensemble but fluctuates on average around zero). The fluctuations are described
by $\var S=\langle S^2\rangle - \langle S\rangle ^2$, which are generally nonzero.
For definiteness, we set $N_1=N_2=N$.
Then, the thermopower fluctuations for both the isothermal and adiabatic cases
are governed by the {\em same} expression:
 \begin{equation}\label{eq_vars}
 \var S=\frac{8\pi^6 k_B^4 \theta^2}{9e^2\delta^2}\frac{1+\delta_{\beta 1}}{(2N+N_3)^4}
\,.
\end{equation}
Our result generalizes the thermopower fluctuations to the case of finite dephasing.
Note that the fluctuations
are not universal and vanish quickly as the mode number increases.
Strikingly, the same functional dependence ($N^{-4}$)
appears in the magnetoasymmetry of the weakly nonlinear conductance \cite{san04}
since it also depends on the energy derivative of the transmission.
The fluctuations are twice larger in the absence of magnetic field.
Importantly, incoherent scattering effects reduce drastically
the fluctuations of $S$ since these have a quantum origin
and the probe introduces decoherence in the system. To leading order,
one has $(\var S-\var{S}^{N_3=0})/\var{S}^{N_3=0}=1-2N_3/N+\mathcal{O}(N_3)^2$.

In the adiabatic case, we have neglected
in $S$ terms that do not contribute to the variance
(e.g., terms that involve three $L$'s are of higher order). Then,
Eq.~\eqref{eq_sad} can be approximated as,
\begin{equation}\label{eq_sapr}
S\simeq \frac{1}{g}
\left[ L_{12}+\frac{G_{32}L_{13}}{G_{31}+G_{32}}
+\frac{G_{13}(G_{31}L_{32}-G_{32}L_{31})}{(G_{31}+G_{32})^2}\right] \,.
\end{equation}
Since for symmetric couplings ($N_1=N_2=N$) the prefactors of $L_{\alpha\beta}$
in Eq. \eqref{eq_sapr} are, on the ensemble average, equal to those of the isothermal case [Eq.~\eqref{eq_S}],
we obtain the same expression for $\var S$. Thus, only asymmetric couplings
($N_1\neq N_2$) could distinguish between the two cases.

{\em Magnetic-field asymmetry}. In a two-terminal conductor, the linear
conductance is an even function of the applied magnetic field $B$.
This fundamental symmetry remains unchanged even after elimination of the probe
coupled to a mesoscopic conductor since $g(B)=g(-B)$. This can be seen by recasting $g$
as $g=G_{11}-G_{13}G_{31}/G_{33}$, which is manifestly symmetric under $B$ reversal.
In contrast, while the two-terminal thermopower
$S=-L_{11}/G_{11}$ is manifestly an even function of $B$, this statement does not hold
in the presence of a probe in the adiabatic case \cite{sai11}.
From Eq.~\eqref{eq_sad} we find the thermopower magnetoasymmetry
$\Phi=S(B)-S(-B)$:
\begin{multline}\label{eq_phi}
\Phi = \left[ G_{33}(L_{13}K_{31}-L_{31}K_{13})+L_{33}(G_{31}K_{13}-G_{13}K_{31})\right.\\
\left. +K_{33}(G_{13}L_{31}-G_{31}L_{13})\right]
/\left[\theta L_{33}(G_{11} L_{33}-G_{31} L_{13})\right. \\
\left.+G_{33} (\theta L_{13} L_{31} +G_{11}K_{33})
-G_{13} (G_{31} K_{33}+\theta L_{31} L_{33})\right]\,,
\end{multline}
which is generally nonzero.
Then, incoherent asymmetry changes the symmetry of the Seebeck coefficient.

We recall that the leading order in a Sommerfeld expansion reads
$K_{\alpha\beta}\simeq -(\pi^2 k_B^2\theta/3e^2) G_{\alpha\beta}$.
Substituting in Eq.~\eqref{eq_phi}, one finds $\Phi=0$.
As a consequence, $B$-asymmetries in the thermopower
are a {\em higher-order} effect. This implies that the size of $\Phi$ will decrease
quickly as temperature decreases (typically, as $\theta^3$). \cite{suppl}
The leading-order Sommerfeld approximation neglects 
terms of order $(k_B \theta/E_F)^4$. \cite{ashcroft}
This factor is rather small in macroscopic metals and
it is thus not necessary to consider higher orders.
But in low dimensional systems the $B$-asymmetry of $S$
should be visible since $E_F$ and $\theta$ can be of the same order.
We note in passing that the conductance $g$ is, in contrast,
always $B$-symmetric independently of the Sommerfeld approximation.
\begin{figure}
\centerline{
\epsfig{file=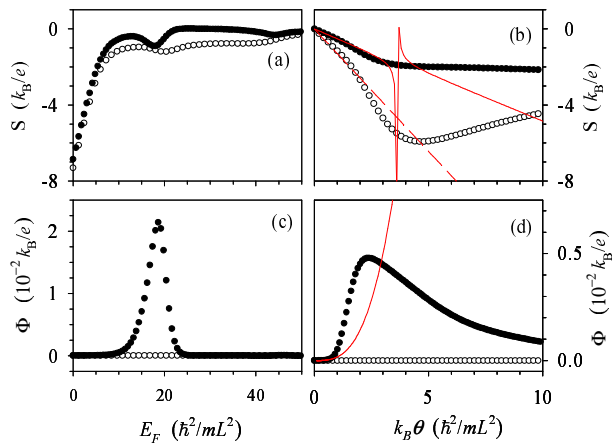,angle=0,width=0.45\textwidth,clip}
}
\caption{
(Color online) Numerical results for the thermopower (a,b) and thermopower 
$B$-asymmetry (c,d) of a microjunction with an asymmetric scatterer [Fig.\ (\ref{fig1}(b)]
as a function of Fermi energy (a,c) and temperature (b,d). 
Full (open) symbols correspond to the presence (absence) of the voltage and thermal probe.
Solid (dashed) lines in (b) are obtained using the Sommerfeld expansion 
to lowest order with (without) the probe.
Note the strong deviation at moderately low temperatures.
Solid line in (d) shows the $\Phi\propto \theta^3$ dependence of the $B$-asymmetry
at low temperature. Parameters: magnetic length $\hbar c/e B = L^2$,
temperature $k_B\theta=\hbar^2/mL^2$ (a,c) and Fermi energy
$E_F=10\hbar^2/mL^2$ (b,d).}
\label{fig2}
\end{figure}

{\em Numerical simulations}. Clearly, for a chaotic cavity
the thermopower magnetoasymmetry vanishes on average. Moreover,
we expect the fluctuations to be exceedingly small since
$\Phi$ depends, to leading order in $\theta$, on the {\em second}
derivative of the transmission. Therefore, it is more natural
to estimate the size of $\Phi$ in a different mesoscopic
system---a microjunction as in Fig.~\ref{fig1}(b).
Thermopowers yielding  0.6~$\mu$V/K have been recently detected
in a similar setup. \cite{mat11} The asymmetric
scatterer deflects the electronic trajectories differently
depending on the $B$ direction. Therefore, we expect an asymmetry
in $S$ when $B$ is inverted.

Our prediction is confirmed by the numerical calculations of Fig.\ \ref{fig2}.
We compute $T_{\alpha\beta}(E)$ using a grid 
discretization of the Schr\"odinger equation and, subsequently,
the response coefficients from Eqs.\ (\ref{eq_rc1})-(\ref{eq_rc4})
with Gauss-Legendre quadratures. \cite{suppl}
Upper panels show that adding the voltage and thermal probe reduces the  
thermopower absolute value as a function of both $E_F$ and $\theta$,
in agreement with our previous results. In Fig.~\ref{fig2}(b) we show that
$S$ deviates from the lowest-order
Sommerfeld expansion as temperature increases both
with and without the probe. We emphasize that
the thermopower is $B$-asymmetric
only if the probe is present, as shown in Figs.~\ref{fig2}(c,d). 
In Fig.~\ref{fig2}d) we plot the asymmetry $\Phi$
as a function of temperature and show that $\Phi$ increases
as $\theta^3$ at low temperature. The maximum value of $\Phi$
depends on the details of the scatterer and the system's parameters.
We find that $\Phi$ can reach values as high as $2$\% for
Fermi energies close to the activation of the 
second mode $E_F\simeq 19.7\hbar^2/mL^2$.

{\em Conclusions}. To summarize, we have investigated
incoherent scattering effects on the thermoelectric transport of a mesoscopic conductor
using a fictitious probe acting as an ideal potentiometer and thermometer.
Our main findings are: (i) a general expression for the
quantum fluctuations of the Seebeck coefficient upon elimination
of the probe valid for both isothermal and adiabatic probes;
(ii) a magnetic-field asymmetry of the thermopower,
requiring both inelastic and dephasing processes,
which becomes apparent only when higher-order terms are
considered in a Sommerfeld expansion.
Odd-in-$B$ thermopowers in Andreev interferometers
have been experimentally observed \cite{cad09}
and theoretically studied. \cite{jac10} We hope that our results will motivate
the experimental detection of asymmetric thermopowers in normal systems.

{\em Acknowledgments}. We thank M. B\"uttiker and R. L\'opez for fruitful discussions.
This work has been supported by the MICINN Grant No.\ FIS2008-00781.

\widetext

\section*{SUPPLEMENTARY MATERIAL}
\subsection{Details on Numerical Calculations}

This section discusses in detail our numerical simulations for a few-mode mesoscopic conductor 
subject to incoherent scattering.
We determine the set of multiterminal transmissions $T_{\alpha\beta}(E)$ using a grid discretization 
of the Schr\"odinger equation and solving the resulting linear equations, varying the boundary conditions
for incidence from each of the leads. The leads are modelled as narrow waveguides
with impenetrable hard walls, forming a T-junction geometry as sketched in Fig.\ \ref{fig1}(b) of the main paper.
We also include an asymmetric triangular scatterer located at the  center of the junction.

\begin{figure}[h]
\centerline{
\epsfig{file=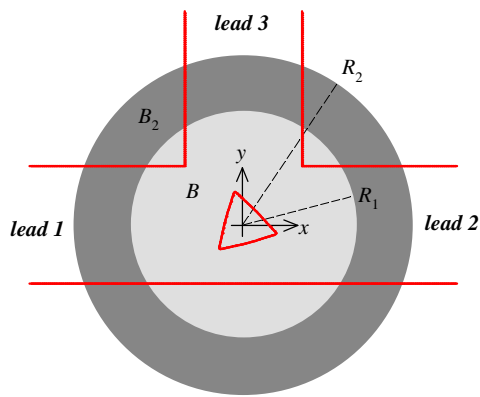,angle=0,width=0.35\textwidth,clip}
}
\caption{(Color online)
Sketch of the magnetic field modelling, with the two regions of radii $R_1$ and $R_2$.
}
\label{fig1_sm}
\end{figure}

Magnetic field effects are modelled as follows. To avoid extended vector-potential fields $\vec{A}(\vec{r}\,)$ 
inside the leads, we consider the localized fields of Fig.\ \ref{fig1_sm}, which might correspond 
to two coaxial coils of radii $R_1$ and $R_2$, perpendicular to the junction and carrying opposite currents. 
In this geometry $\vec{B}(\vec{r}\,)=B(r)\hat{u}_z$
and $\vec{A}(\vec{r}\,)=A(r)\hat{u}_\phi$, where $\hat{u}_z$ and $\hat{u}_\phi$ are the Cartesian 
and the azimuthal unit vectors, respectively, while $r$ is the distance from the 
junction center (Fig.\ \ref{fig1_sm}). The field strengths read 
\begin{equation}
\begin{array}{c|c|c}
 & B(r) & A(r) \\
\hline
r < R_1 & B & \frac{B}{2}r \rule{0cm}{0.5cm} \\
R_1<r<R_2 & -B\frac{R_1^2}{R_2^2-R_1^2} &  \frac{B}{2}\frac{R_1^2R_2^2}{R_2^2-R_1^2}
\left(
\frac{1}{r}-
\frac{r}{R_2^2}
\right) \rule{0cm}{0.5cm}\\
r > R_2 \rule{0cm}{0.5cm} & 0 & 0 
\end{array}\; ,
\end{equation}
Notice that, as required, not only $B(r)$, but also $A(r)$, vanishes in the leads when $r>R_2$. 
Therefore, our results arise purely from magnetic effects that occur within the junction only.

The central scatterer is modelled as a smooth two-dimensional barrier of height $V_0$
with a triangular shape in the plane. More
specifically, the scatterer's potential barrier in polar coordinates is given by
\begin{equation}
V(r,\theta)=\frac{V_0}{1+e^{(r-R_T(\theta))/\sigma_T}}\; ,
\end{equation}
where $R_T(\theta)$ is a piecewise function characterizing a triangle of side $L_T$ and oriented along
a specific angle $\theta_T$, and $\sigma_T$ a small diffusivity.

\begin{figure}[h]
\centerline{
\epsfig{file=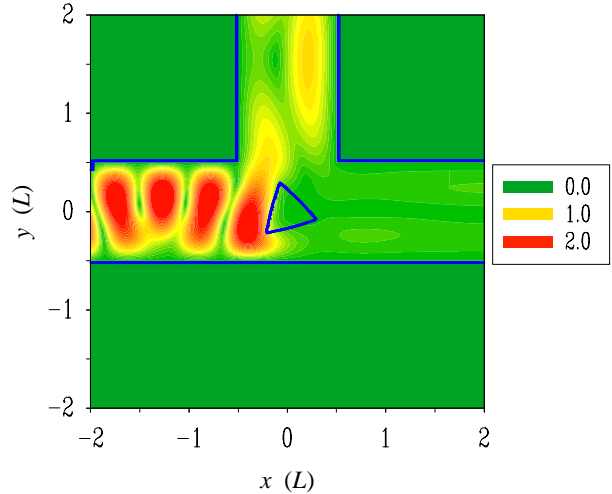,angle=0,width=0.45\textwidth,clip}
}
\caption{(Color online)
Density distribution, $|\Psi(\vec{r}\,)|^2$, with $\Psi(\vec{r}\,)$ the solution of the
Schr\"odinger equation for incidence from terminal 1 (left).
The density is given in units of $L^{-2}$, where $L$ is the lateral extension of
the leads. The position of the leads and the scatterer is indicated by solid lines.
The incident energy is $E=30 \hbar^2/mL^2$.
Parameters of the magnetic field: $\hbar c/eB=L^2$, $R_1=L$, $R_2=1.5L$.
Parameters of the repulsive scatterer: $V_0=200\hbar^2/mL^2$, $L_T=0.5L$,
$\sigma_T=0.05L$, $\theta_T=15^{\rm o}$. 
}
\label{fig2_sm}
\end{figure}

The numerical calculation of the kinetic terms including vector potential contributions is made using 
the Peierls substitution for lattice calculations. This is very convenient since it fulfills gauge invariance 
by construction. The practical 
implementation is as follows.
We denote with $t_{ij}$ the finite-difference matrix associated with the 
kinetic energy on the grid. Thus, the operator acting on function $\Psi$ at grid point $i$ 
is given by the following sum over neighboring points $j$:
\begin{equation}
\left[\frac{(-i\hbar\nabla)^2}{2m} \Psi\right]_i = \sum_j{t}_{ij}\,\Psi_j\; .
\end{equation}
The Peierls substitution for the vector potential terms amounts to including the following field-dependent phases
in the sum
\begin{equation}
\left[\frac{1}{2m}
\left(-i\hbar\nabla+\frac{e}{c}\vec{A}\right)^2 \Psi\right]_i = \sum_j{t}_{ij}\, 
e^{{\frac{ie}{\hbar c}\frac{\vec{A}_i+\vec{A}_j}{2}\cdot(\vec{r}_i-\vec{r}_j)}}\,
\Psi_j\; .
\end{equation}
As an example, Fig.\ \ref{fig2_sm} displays the probability density in the presence of scatterer and 
magnetic field in the case of incidence from the left (terminal 1). 

Once the wave function for each incidence 
condition is known, it is an straightforward task to determine the quantum transmissions $T_{\alpha\beta}$ to be used 
in the evaluation of the linear transport coefficients, Eqs.\ (4)-(7) of the paper.
These integrals are performed using a controlled numerical method, with the required 
precision ($10^{-5}$) achieved by successive partitioning of the integration domain 
in an increasing number of zones and computing each of them with a Gauss-Legendre method. 
In practice, the transmissions
$T_{\alpha\beta}(E)$ for arbitrary energy $E$ are obtained by interpolating or extrapolating 
on a table of closely spaced precomputed values.
Special care is taken for $E$ values close to the transverse mode energies to keep 
the discontinuities at these energies of the transmissions.

\section{Low temperature dependence}

In this section, we show that at low temperature the thermopower asymmetry has a $\theta^3$ dependence. We
present our analysis, both numerical and analytical, of this limit.
We recall that the thermopower $B$-asymmetry $\Phi=S(B)-S(-B)$ reads,
\begin{equation}\label{eq_phi_sm}
\Phi = \frac{G_{33}(L_{13}K_{31}-L_{31}K_{13})+L_{33}(G_{31}K_{13}-G_{13}K_{31})
+K_{33}(G_{13}L_{31}-G_{31}L_{13})
}{\theta L_{33}(G_{11} L_{33}-G_{31} L_{13})
+G_{33} (\theta L_{13} L_{31} +G_{11}K_{33})
-G_{13} (G_{31} K_{33}+\theta L_{31} L_{33})}\,,
\end{equation}
where the transport coefficients are given by
\begin{align}
G_{\alpha\beta} &=g_V (N_\alpha\delta_{\alpha\beta}-T_{\alpha\beta})\,,\label{eq_tr1}\\
L_{\alpha\beta} &=g_\theta T_{\alpha\beta}'\,,\label{eq_tr2}\\
K_{\alpha\beta} &=-g_\theta (N_\alpha\delta_{\alpha\beta}-T_{\alpha\beta})/e\,,\label{eq_tr3}
\end{align}
to leading order in the Sommerfeld expansion. Higher orders are $\mathcal{O}(\theta)^3$.
It follows from Eqs.~\eqref{eq_tr1} and~\eqref{eq_tr3} that
$K_{\alpha\beta}=-(\pi^2 k_B^2/3e^2)\theta G_{\alpha\beta}$.
Substituting this result in Eq.~\eqref{eq_phi} one obtains $\Phi=0$.
Therefore, nonzero magnetoasymmetries arise only if we carry out a Sommerfeld expansion
for $K_{\alpha\beta}$ to higher order. We find,
\begin{equation}
K_{\alpha\beta}=-\frac{\pi^2 k_B^2}{3e^2}\theta G_{\alpha\beta}
+\frac{7\pi^4 k_B^4}{15h}\theta ^3 T_{\alpha\beta}''+\mathcal{O}(\theta)^5 \,.
\end{equation}
Since $L_{\alpha\beta}$ is $\mathcal{O}(\theta)$, the numerator in the right-hand side
of Eq.~\eqref{eq_phi_sm} is $\mathcal{O}(\theta)^4$, whereas the denominator is $\mathcal{O}(\theta)$.
Then, the leading order of $\Phi$ is $\mathcal{O}(\theta)^3$.

\begin{figure}[h]
\centerline{
\epsfig{file=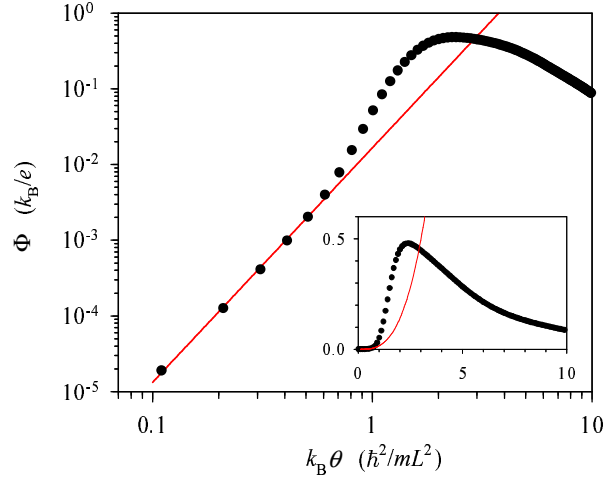,angle=0,width=0.45\textwidth,clip}
}
\caption{(Color online) Low temperature fit  of the thermopower asymmetry, $\Phi\propto\theta^3$.
Logarithmic scale is used in the main plot
to highlight the low temperature region. For comparison, the inset shows the same data in linear scale.
We use the same parameters as in Fig.\ 3(d) of the main paper: 
$E=10 \hbar^2/mL^2$, 
$\hbar c/eB=L^2$,
$R_1=L$, $R_2=1.5L$,
$V_0=200\hbar^2/mL^2$, $L_T=0.5L$,
$\sigma_T=0.05L$ and $\theta_T=15^{\rm o}$.
}
\label{fig3_sm}
\end{figure}

Our numerical results are shown in Fig.\ \ref{fig3_sm}, which displays 
a fit of $\Phi$ versus $\theta$ (dots) to a cubic-law behavior (line).
The behavior predicted above
is in agreement with the numerical data at low temperature. Obviously, higher-order contributions
start to dominate as temperature increases. For the asymmetric microjunction,
we estimate that the thermopower asymmetry deviates from the $\theta^3$ law
at $k_B\theta\gtrsim\hbar^2/mL^2$.

\end{document}